\def\year{2020}\relax
\documentclass[letterpaper]{article} 
\usepackage{aaai20}  
\usepackage{times}  
\usepackage{helvet} 
\usepackage{courier}  
\usepackage[hyphens]{url}  
\usepackage{graphicx} 
\usepackage{graphicx}  
\usepackage{multirow}

\tiny
\title{I Know Where You Are Coming From: \\On the Impact of Social Media Sources on AI Model Performance(Student Abstract)}

\author{ \Large \textbf{Yang Qi, Farseev Aleksandr, Filchenkov Andrey}\\ 
 ITMO University\\ 
49 Kronverksky Pr., St. Peterburg, 197101, +78122329704\\
yangqi@itmo.ru, farseev@itmo.ru, afilchenkov@itmo.ru 
}

\begin{document}

\maketitle

\begin{abstract}
Nowadays, social networks play a crucial role in human everyday life and no longer purely associated with spare time spending. In fact, instant communication with friends and colleagues has become an essential component of our daily interaction giving a raise of multiple new social network types emergence. By participating in such networks, individuals generate a multitude of data points that describe their activities from different perspectives and, for example, can be further used for applications such as personalized recommendation or user profiling. However, the impact of the different social media networks on machine learning model performance has not been studied comprehensively yet. Particularly, the literature on modeling multi-modal data from multiple social networks is relatively sparse, which had inspired us to take a deeper dive into the topic in this preliminary study. Specifically, in this work, we will study the performance of different machine learning models when being learned on multi-modal data from different social networks. Our initial experimental results reveal that social network choice impacts the performance and the proper selection of data source is crucial.
\end{abstract}

In recent years, numerous new types of social media platforms have emerged. They bridge human interactions across the globe and serve as an important medium for daily communication. To satisfy various human communication demands, many of such social networks have been historically adopted to serve different purposes. For example, in Japan Twitter is often used for formal interaction, while Instagram is conventionally treated as a visual co-sharing venue for younger adults. At the same time, one can observe that the way people are using conventional social networks, such as Facebook, overtime shifts towards private messaging with no much public exposure.

By participating in social media platforms, individuals generate a multitude of multi-modal content that describes their activities from diverse perspectives. In turn, such content can be further used for different applications, such as personalized recommendation or automatic user profiling. These applications are essential for a wide range of modern value-driven industry and research scenarios. For example, Personality Traits, that reflect human emotional characteristics and mental status, can be efficiently utilized in marketing, HR, and public spaces. At the same time, such knowledge could also empower various research discoveries, driving a better understanding of the reasons behind certain human behavior.

From the past works, it is evident that Machine Learning models are able to achieve higher performance when being trained on the data of multiple modalities when such multi-view data is modeled in a mutually-consistent and comprehensive fashion prior or during~\cite{Farseev} the learning process. However, the impact of the different data sources (social networks) on machine learning performance have not been comprehensively studied yet, and thus require further investigation.

Inspired by the research gap above, in this study, we are aiming at further exploration of the impact of social media usage patterns and the diversity of the content to the machine learning model performance. Particularly, we investigate the impact of the social media network choice on the performance of the multi-view machine learning models when being trained to predict human personality based on public multi-modal social media content.

Personality profiling from multi-source multi-view data is a challenging problem. First, it requires a proper Data Gathering approach that is able to gather multi-view data from different social networks equipped with the sensitive personality ground-truth label, which is hard due to the privacy restrictions and the lack of the public knowledge about the personality of social network users. Second, the integration and further modeling of such unstructured multi-view data from multiple sources in a mutually-consistent and comprehensive manner is a well-known challenging machine learning problem alone.

Inspired by the challenges and the research gap above, in this study we are asking the following research question: \textbf{what is the role of different data sources and modalities when learning from different Social Networks?}

Following the idea of choosing social networks of diverse and different user usage patterns, we've picked up Facebook and Twitter as the two main data sources. Twitter has been chosen as it is the World's largest English-speaking micro-blogging service where users share brief daily updates. Facebook was chosen as the World's largest inter-person communication medium that is often used for private discussions. We've harvested the public multi-view user-generated data (images and text) and  Personality ground truth (Extrovert V.S. Introvert) from both social networks. The ground truth was extracted from users' public shares. For every collected profile, we've also downloaded all the public posts uploaded by the users to assemble our final dataset. The numbers of data points with respect to each data modality/source and ground truth labels are indicated in the Table~\ref{tab:dataset-statistics}.

\begin{table}[t]
\caption{Dataset Statistics}\smallskip
\centering
\resizebox{0.65\columnwidth}{!}{
\smallskip\begin{tabular}{l|l|l|}
\cline{1-3}
\multicolumn{1}{|l|}{number} & Facebook & Twitter \\ \hline
\multicolumn{1}{|l|}{\# users}  & 10252    & 2795    \\ \hline
\multicolumn{1}{|l|}{\# posts}  & 2255734  & 2416468 \\ \hline
\multicolumn{1}{|l|}{\# images} & 466822   & 231627  \\ \hline
\multicolumn{1}{|l|}{\# Extroverts} & 5362   & 1677  \\ \hline
\multicolumn{1}{|l|}{\# Introverts} & 4890   & 1168  \\ \hline

\end{tabular}}
\label{tab:dataset-statistics}

\end{table}

To model the impact of the data source choice on the performance of personality profiling, we've first represented multi-modal data in a form of feature vectors as follows:

\begin{itemize}
\item \textbf{Textual data} was converted into Tf-Idf space and the latent semantic analysis (LSA)~\cite{LSA} was applied thereafter. The approach was chosen due to its effectiveness reported in the previous studies for similar applications of user profiling~\cite{Daneshvar2018}. The number of dimensions was set to $200$, which was found via a grid search and allowed for achieving the best prediction performance.

\item \textbf{Image data} was represented as image concept vector, inferred via pre-trained ResNet101 architecture on the ImageNet dataset~\cite{he2016deep}. Similarly to the text modality, we've reduced the feature vector dimension to $200$ by applying LSA, which has further helped to boost the prediction performance.
\end{itemize}

To test the hypothesis and answer our research question, we've selected the following three machine learning approaches that have been previously shown to perform well for the task of User Profiling in the recent PAN evaluations: Light Gradient Boosting Machine (LGBM), Linear Regression (LR) and  Support Vector Machine with the linear kernel (Linear SVC). The above three models were trained based on the multi-view data to perform personality classification from different modalities independently, as well as combined together. The prediction performance was evaluated in terms of F-score via 10-fold Cross-Validation.

\begin{table}[t]
\caption{Multi-View Model Performance Comparison}\smallskip
\centering
\resizebox{0.65\columnwidth}{!}{
\begin{tabular}{|c|c|c|c|}

\hline
\multirow{2}{*}{Modality}   & \multirow{2}{*}{Classifier} & \multicolumn{2}{c|}{EI} \\ \cline{3-4} 
                            &                             & Facebook    & Twitter   \\ \hline
\multirow{3}{*}{Text Only}  & LR                          & 0.67        & 0.66      \\ \cline{2-2}
                            & SVC                         & 0.65        & 0.64      \\ \cline{2-2}
                            & LGBM                        & 0.64        & 0.62      \\ \hline
\multirow{3}{*}{Image Only} & LR                          & 0.63        & 0.55      \\ \cline{2-2}
                            & SVC                         & 0.61        & 0.59      \\ \cline{2-2}
                            & LGBM                        & 0.65        & 0.57      \\ \hline
\multirow{3}{*}{Text+Image} & LR                          & 0.61        & 0.65      \\ \cline{2-2}
                            & SVC                         & 0.69        & 0.61      \\ \cline{2-2}
                            & LGBM                        & 0.64        & 0.64      \\ \hline
\end{tabular}}
\label{tab:result}
\end{table}

The evaluation results are presented in Table~\ref{tab:result}. From the table, it can be seen that all three models have achieved similar performance when being trained on a single text modality with a slight LR and SVC improvement over the LGBM model. The finding demonstrates that, even though the social network usage patterns are drastically different in Twitter and Facebook, the \textbf{textual modality alone has similar predictive performance} for different social networks and tent to be handled better by the linear models. The above can be explained by the simplicity of the text data as compared to more complex and diverse visual data representations.

At the same time, one can notice the significant differences of the models' performance when being trained solely based on the image modality. Specifically, it can be seen that all the models were able to benefit from the visual modality more on Facebook than on Twitter. The latter demonstrates that \textbf{visual data source choice significantly impacts the model performance} and that the Facebook visual data is of higher predictive power as compared to Twitter. The finding can be explained by the diversity of the personal visual content exposed on Facebook: being utilized in machine learning, such content is able to help the models in predicting personality more accurately as compared to Twitter data, where the message size is limited and the communication style is less visual and more concise. The above findings also align well with the feature weight analysis, showing that, \textbf{the weights trained from visual modality are significantly different for Facebook and Twitter.}

Finally, it can also be seen that the models trained based on multi-modal data significantly outperform the single-modal cases with the best performance achieved by the SVC for Facebook. The above conforms well with our previous findings showing that \textbf{multi-source data utilization is beneficial for Social Multimedia Learning applications while visual data from Facebook helps in improving performance of both single-modal and multi-modal models}.

In our future works we plan to further explore the impact of different data sources to the performance of fusion models acting in the cross-lingual and cross-regional environments.

\section{Acknowledgments}
This research was funded by the Government of Russian Federation, Grant 08-08.

\fontsize{9.0pt}{10.0pt}
\selectfont
\bibliography{mybibliography.bib}
\bibliographystyle{aaai}

\end{document}